Mini-Review

# Photoproduction of $\pi^0$-Mesons from the Nucleon and Deuteron

Eed Mohamed Darwish[1,2*], Amel Hemmdan[3], and Nesreen T. El-Shamy[1,4]

1. Physics Department, Faculty of Science, Taibah University, Madina 41477, Saudi Arabia
2. Physics Department, Faculty of Science, Sohag University, Sohag 82524, Egypt
3. Physics Department, Faculty of Science, Aswan University, Aswan 81528, Egypt
4. Physics Department, Faculty of Girls, Ain Shams University, Heliopolis, Cairo 11566, Egypt

[*]E-mail: edarwish@taibahu.edu.sa

## Abstract

The photoproduction of $\pi^0$-mesons from the nucleon and deuteron has been studied for incidents of photon energies up to 1.5 GeV. By using the MAID-2007 model for the process on the nucleon, we predict results for the unpolarized and helicity-dependent total cross sections of the semi-exclusive reaction $\gamma d \to \pi^0 X$ ($X=np+d$) with the inclusion of rescattering effects. We find that rescattering effects yield a substantially large contribution. The extracted results are compared with the available experimental data and a satisfactory agreement is obtained. In addition, the contribution of $\gamma d \to \pi^0 X$ ($X=np+d$) to the finite GDH integral has been evaluated by explicit integration up to 1.5 GeV and a total value of 256.96 µb has been obtained. Convergence of the GDH integral has been reached.

## Abstrak

**Fotoproduksi Meson-π0 dari Nukleon dan Deuteron.** Fotoproduksi meson-$\pi^0$ dari nukleon dan deuteron telah dipelajari untuk energi-energi foton datang hingga 1.5 GeV. Dengan menggunakan model MAID-2007 untuk proses produksi pada nukleon, kami meramalkan hasil-hasil perhitungan penampang lintang total tidak-terpolarisasi dan penampang lintang total bergantung-helisitas dari reaksi semi-eksklusif $\gamma d \to \pi^0 X$ ($X=np+d$) dengan memasukkan efek *rescattering*. Hasil studi memperlihatkan bahwa efek *rescattering* memberikan kontribusi sangat besar. Perbandingan hasil-hasil ekstraksi dengan data eksperimen yang tersedia memperlihatkan kecocokan yang memuaskan. Sebagai tambahan, kontribusi dari proses $\gamma d \to \pi^0 X$ ($X=np+d$) pada integral GDH berhingga telah dihitung dengan menggunakan integrasi eksplisit hingga 1.5 GeV dan nilai total sebesar 256.96 µb telah diperoleh. Konvergensi dari integral GDH telah tercapai.

*Keywords: few-body systems, meson production, photoproduction reactions, spin observables*

## Introduction

The photoproduction of neutral mesons plays a special role for the study of resonance contributions, since background contributions like meson pole terms or Kroll-Rudermann terms are strongly suppressed for the proton, but not for the neutron due to the weak coupling of the photon to neutral mesons. The photoproduction of mesons from the free proton has been developed as the most important tool for the study of the excitation spectrum of the nucleon. A huge experimental effort has been performed at several electron accelerators (MAMI, ELSA, LEGS, ESRF, JLab, MAX-Lab) to study not only unpolarized cross sections, but also polarization observables for many different final states [1,2].

Photoproduction from the proton alone tells us nothing about the isospin dependence of the electromagnetic interactions. It is thus desirable to also investigate the same reactions on the neutron target, which provide the only access to the isospin decomposition of the electromagnetic excitation amplitudes. But due to the non-availability of free neutron targets, one must rely on meson photoproduction from light nuclei. In principle, two possible mechanisms need to be learned about the isospin structure of the photoexcitation amplitudes: breakup and coherent. In the breakup mechanism, the nucleon is not at rest; it is in motion, the so-called Fermi motion. Thus, in the case of the breakup mechanism, the reaction is quasifree, and the Fermi motion and possible final-state interaction (FSI) effects have to be taken into account. In contrast, the coherent mechanism is simpler as the nucleus in the initial and final states is identical. The small binding energy and the comparatively well understood nuclear structure single out the deuteron as exceptionally important target nucleus. Furthermore, the photoproduction of mesons from the deuteron bridges the gap between the elementary reaction on the free proton and the photoproduction from heavier nuclei, which is very interesting for the study of the behavior of nucleon resonances in the nuclear medium.

Some of the recent progress in the physics of $\pi^0$-photoproduction from the deuteron is reviewed from a theoretical perspective. In recent years, several studies (see e.g. [3-8] and references therein) have been performed in coherent and incoherent $\pi^0$-photoproduction from the deuteron. Despite these considerable theoretical efforts, a satisfactory overall description of the experimental data has yet to be found [1,2]. It was found that the photoproduction cross sections of $\gamma d \rightarrow \pi^0 np$ are overestimated. In addition, the shape of the $\gamma d \rightarrow \pi^0 d$ cross sections is very poorly reproduced, especially at photon energies near η-threshold. Turning to polarization observables, the helicity-dependent cross sections from the semi-exclusive $\gamma d \rightarrow \pi^0 X$ ($X=np+d$) reaction are also not in good agreement with experimental data.

The ultimate goal of this work is to report on a theoretical prediction for the semi-exclusive reaction $\gamma d \rightarrow \pi^0 X$ ($X=np+d$) for incident photon energy up to 1.5 GeV by using the isobar MAID-2007 model [9] for the elementary $\gamma N \rightarrow \pi^0 N$ process. The influence of rescattering effects on the unpolarized and helicity-dependent total cross sections as well as on the Gerasimov-Drell-Hearn (GDH) sum rule [10] will be investigated. We will also compare the predicted results with the available experimental data. The calculation is of high theoretical interest, because it provides an important test of our understanding of the $\pi^0 np$ and $\pi^0 d$ dynamics, which is a prerequisite for reliable extraction of the elementary $\gamma n \rightarrow \pi^0 n$ amplitude.

**Methods**

In this section, we briefly outline the model used to evaluate the transition amplitudes for coherent and incoherent $\pi^0$-photoproduction from the deuteron. Formal details of this model were described in Refs. [3,7], to which the reader is referred for more information.

Within our model for the $\gamma d \rightarrow \pi^0 np$ reaction, the scattering amplitude is treated in such a way that besides the pure impulse approximation (IA$_{incoh}$), the complete rescattering by FSI within each of the $np$ and $\pi^0 N$ two-body subsystems is included. Therefore, the total transition matrix element reads in this approximation

$$M^{incoh} = M^{IA}_{incoh} + M^{np} + M^{\pi^0 N}. \qquad (1)$$

A graphical representation of the scattering matrix is shown in Fig. 1. Diagram (a) is often referred to as IA$_{incoh}$ or the pole diagram. Apart from that, $np$-FSI [diagram (b)] and $\pi^0 N$-FSI [diagrams (c) and (d)] have been considered.

The $\gamma d \rightarrow \pi^0 np$ amplitude is evaluated by taking the isobar MAID-2007 model [9] for the elementary operator, allowing one to extend the calculation up to photon energies of 1.5 GeV. Moreover, FSI effects are perturbatively taken into account up to the first order in the corresponding $np$- and $\pi^0 N$-scattering amplitudes. The necessary half-off-shell two-body scattering matrices were obtained from separable representations of realistic $np$- and $\pi^0 N$-interactions, which gave good descriptions of the corresponding phase shifts [11]. For $np$-rescattering, we have included all partial waves with total angular momentum $J \leq 3$ and for $\pi^0 N$-rescattering $S$- through $D$-waves. For further details with respect to the rescattering terms, we refer the reader to our previous work [3].

For coherent $\pi^0$-photoproduction from the deuteron, the calculation of the transition amplitude follows the standard recipe in which the pure impulse approximation (IA$_{coh}$) [diagram (a) in Fig. 2] is treated as the first basic approximation, and the contributions from the intermediate two-body process with $\pi N$- and $\eta N$-rescattering (FOR) and the higher-order terms in the multiple scattering series (TBM) effects are taken into account in the form of additional terms, which are presented in the diagrams (b) and (c). In order to avoid double counting, diagrams (d) and (e) in Fig. 2 are removed from the three-body amplitude since they possess the same topology as that already included in diagrams (a) and (b). For the calculations of diagrams (c), (d), and (e), only the $s$-wave state $^1S_0$ ($J^\pi, T = 0^-, 1$) is considered in this work. In addition, the $T_{S11}$ amplitude of photoproduction of the $S_{11}(1535)$ resonance from the deuteron (Fig. 3) is calculated only in the $s$-wave state $^1S_0$ ($J^\pi, T = 0^-, 1$).

The three contributions in diagram (a), diagrams (a)+(b), and diagrams (a)+(b)+(c) in Fig. 2 present three different

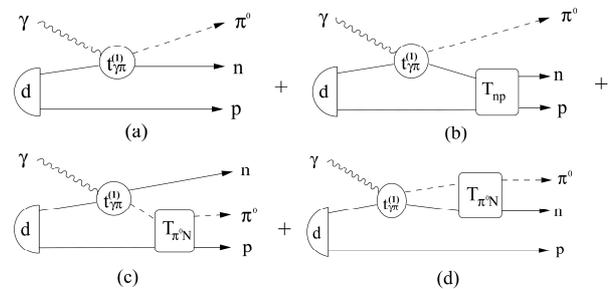

**Figure 1.** Considered Diagrams for the $\gamma d \rightarrow \pi^0 np$ Amplitude. Diagram (a) Corresponds to IA$_{incoh}$, (b) Incorporation of $np$-FSI, (c) and (d) Incorporation of $\pi^0 N$-FSI

levels of approximation to the reaction amplitude. The total scattering $M^{coh}$-matrix reads in this approximation

$$M^{coh} = M^{IA}_{coh} + M^{FOR} + M^{TBM} . \quad (2)$$

The procedure, which we have used to calculate the three terms in (2), is described in detail in Ref. [7].

As in the case of incoherent reaction, the $\gamma d \rightarrow \pi^0 d$ amplitude is evaluated by taking the MAID-2007 model [9] for the elementary operator. For the various hadronic and electromagnetic two-body reactions included in the treatment of the rescattering diagrams, only the $S_{11}(1535)$ resonance was taken into account. Furthermore, the $\pi$- and $\eta$-photo-production on the nucleon as well as their interactions with nucleons were assumed to be proceeded exclusively via the extraction of the $S_{11}(1535)$-resonance. The three-body problem of the intermediate $\eta NN$

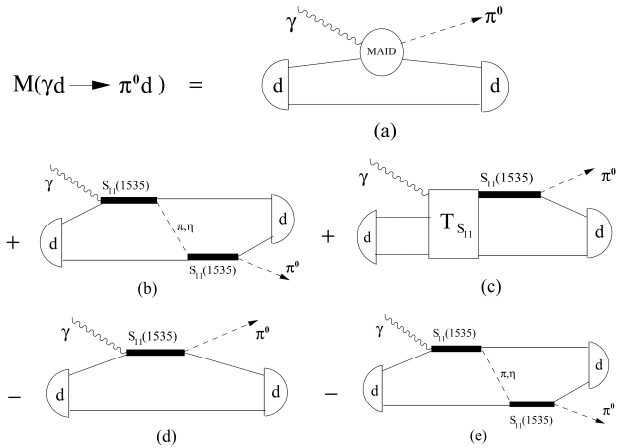

**Figure 2.** The $\gamma d \rightarrow \pi^0 d$ Amplitude Includes Rescattering Effects in the Intermediate state (a) Impulse Approximation ($IA_{coh}$), (b) Two-step process with Intermediate $\pi N$- and $\eta N$-rescattering (FOR), and (c) Contribution from Three-body Dynamics (TBM). In Order to Avoid Double Counting, Diagrams (d) and (e) are Removed from the 3-body Amplitude Since they Possess the same Topology as the ones Already Included in Diagrams (a) and (b)

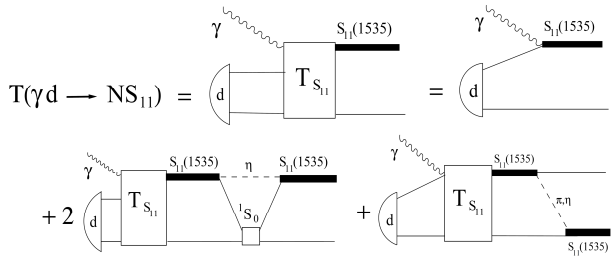

**Figure 3.** Diagrammatic Representation for the $T_{S11}$ Amplitude of Photoproduction of the $S_{11}(1535)$-resonance from the Deuteron (as Shown in Diagram (c) of Fig. 2) Calculated Only in the s-wave State $^1S_0$ ($J^\pi$, $T = 0^-$, 1)

interaction has been solved by using a separable representation of the driving two-body interaction in the $\pi N$, $\eta N$, and $NN$ subsystems. The corresponding amplitudes for $\pi N$ and $\eta N$ scattering are given by an isobar ansatz from Ref. [8]. For the $NN$ subsystem, the separable representation of the Bonn potential [12] has been used.

## Results and Discussion

We begin the discussion of the results with the unpolarized total cross section for the semi-exclusive reaction $\gamma d \rightarrow \pi^0 X$ ($X=np+d$) displayed in Fig. 4 as a function of photon energy in the laboratory system in the IA alone (dotted curve) and with the inclusion of rescattering effects (solid curve). In addition, we have plotted the corresponding elementary cross section for the $\gamma N \rightarrow \pi^0 N$ reaction (dashed curve) for comparison. For this purpose, the MAID-2007 model [9] as parametrized in terms of invariant amplitudes defined in Ref. [13] is taken. It is readily noted that rescattering effects are quite large, especially around the $\Delta(1232)$-resonance region. The main contribution comes from $np$-rescattering, while $\pi^0 N$-rescattering is almost negligible. FSI effects significantly reduce the total cross section up to 500 MeV from its IA value. As was mentioned in Ref. [6], these effects are mainly due to the elimination of the spurious coherent contribution in the IA. With respect to the elementary cross section, a sizable reduction for incoherent $\pi^0$-photoproduction from the deuteron is found, because part of the strength goes to the coherent channel $\gamma d \rightarrow \pi^0 d$. The spurious admixture of the coherent channel in the IA cross section is also discussed in Refs. [6,14]. It is also noted that the resulting total cross section for the elementary reaction shows some reduction compared with the deuteron cross section that includes FSI effects, because of the Fermi motion.

The results of the helicity-dependent total cross section difference $\sigma^P - \sigma^A$ for coherent and incoherent $\pi^0$-photoproduction channels from the deuteron in IA only and with the inclusion of rescattering effects are shown in the upper left panel of Figure 5. The upper right panel shows the results for the semi-exclusive reaction $\gamma d \rightarrow \pi^0 X$ ($X=np+d$) together with the corresponding results for the $\gamma N \rightarrow \pi^0 N$ reaction channels. A positive contribution to the deuteron spin asymmetry from the $\Delta(1232)$-resonance is obtained. We found that rescattering effects are quite sizeable for the $\gamma d \rightarrow \pi^0 np$ channel due to the non-orthogonality of the final state plane wave in $IA_{incoh}$ to the deuteron bound state wave function. Comparing the deuteron and nucleon results, a significant difference is obtained, especially in the peak position (see the difference between the solid and dashed curves in the upper right panel of Figure 5). This difference comes from the Fermi motion and leads to a big increase in the elementary helicity-dependent cross section difference $\sigma^P - \sigma^A$.

The lower left panel of Figure 5 exhibits the separate contributions of the $\gamma d \to \pi^0 np$ and $\gamma d \to \pi^0 d$ channels to the finite GDH integral for the deuteron as a function of the upper integration limit. The lower right panel displays the corresponding results for the semi-exclusive reaction $\gamma d \to \pi^0 X$ ($X=np+d$) in IA (dotted curve) and with the inclusion of rescattering effects (solid curve) in comparison with the results of the free nucleon by using MAID-2007 model (dashed curve). In addition to the IA, a significant influence of rescattering effects is found. The finite GDH integral of the nucleon and deuteron also shows a big difference between both results, because of the Fermi motion. It is obvious that convergence of the GDH integral is reached. Such convergence has also been reached in Ref. [4].

Table 1 displays the difference between $\pi^0$-photoproduction from the nucleon and deuteron, where separate contributions for the various channels to the finite GDH integral up to 1.5 GeV are listed. We found a large contribution to the deuteron GDH integral stems from $\gamma d \to \pi^0 np$ channel. For the total contribution of the semi-exclusive reaction $\gamma d \to \pi^0 X$ ($X=np+d$), a value of 256.96 µb is obtained by explicit integration up to 1.5 GeV, while in Ref. [4] the value 315.75 µb was obtained. The difference between both values seems likely to be due to the neglected contributions from two-body mechanisms.

Figure 6 displays the helicity-dependent total cross section difference $\sigma^P - \sigma^A$ for the semi-exclusive reaction $\gamma d \to \pi^0 X$ ($X=np+d$) with the inclusion of rescattering effects as a function of photon lab-energy up to 450 MeV. A comparison with the experimental data from MAMI [2] and with the results of the free nucleon is also shown. It can be seen that our predictions with the inclusion of rescattering effects (solid curve) adequately reproduce the experimental data. We would like to emphasize that the values of spin asymmetry in the $\Delta(1232)$-resonance region for the $\pi^0 np$ channel are reduced by about 40% when rescattering effects are added to the pure quasifree mechanisms. It is also obvious that the agreement between our predictions and the experimental data is satisfactory, and it is much better than in the unpolarized case (see Ref. [3]). This might be an indication that the reaction mechanisms that are not taken into account in the cross section calculation do not have a pronounced helicity-dependence. The nucleon results are not significantly different from the deuteron results. Only in the peak position is a noticeable difference obtained.

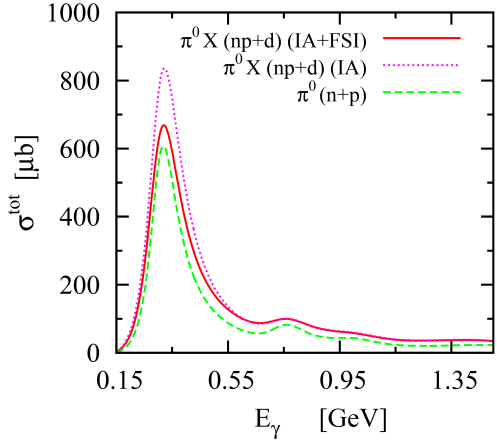

**Figure 4.** Total $\pi^0$-photoproduction Cross Section from the Nucleon and Deuteron. The Deuteron Results for $\gamma d \to \pi^0 X$ ($X = np+d$) are in IA (Dotted Curve) and with the Inclusion of Rescattering Effects (Solid Curve). The Nucleon Results (Dashed Curve) are the Sum of the Proton and Neutron Targets that are Predicted Using MAID-2007 Model [9]

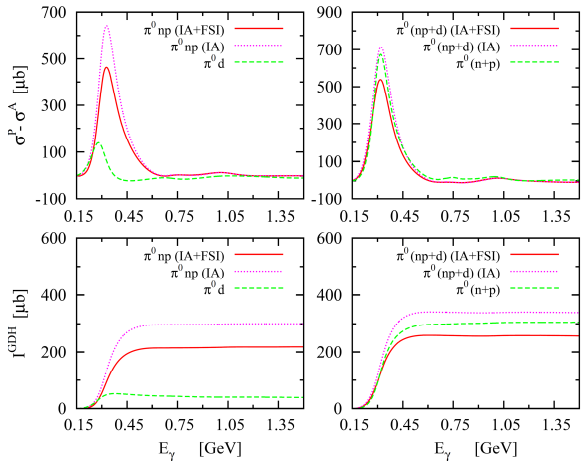

**Figure 5.** Upper Part: the Helicity-dependent Total Cross Section Difference $\sigma^P - \sigma^A$ of $\pi^0$-photoproduction from the Nucleon and Deuteron. Lower Part: the Corresponding Contribution of Various Channels to the Finite GDH Integral as a Function of the Upper Integration Limit. The Nucleon Results are the Sum of the Proton and Neutron Targets that are Predicted Using MAID-2007 Model [9]

**Table 1.** Contribution of $\pi^0$-photo-production Channels on the Nucleon and Deuteron to the Finite GDH Integral up to 1.5 GeV in *µb*. The Nucleon Results are Predicted Using MAID-2007 Model [9]

| nucleon | $\pi^0 p$ | $\pi^0 n$ | Total | Ref. [4] |
|---------|-----------|-----------|--------|----------|
|         | 154.14    | 151.30    | 305.44 | 306.44   |
| deuteron | $\pi^0 np$ | $\pi^0 d$ | total | Ref. [4] |
|         | 218.21    | 38.75     | 256.96 | 315.75   |

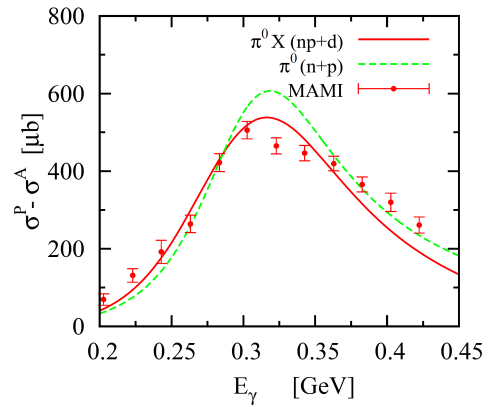

**Figure 6.** The Helicity-dependent Total Cross Section Difference $\sigma^P - \sigma^A$ of $\gamma d \to \pi^0 X$ ($X=np+d$) with the Inclusion of Rescattering Effects Compared with the Data from MAMI [2] and the Results of the Free Nucleon by Using MAID-2007 Model [9]

## Conclusions

We have presented a selection of results obtained in a model used to describe the coherent and incoherent $\pi^0$-photoproduction from the deuteron via the semi-exclusive reaction $\gamma d \rightarrow \pi^0 X$ ($X=np+d$) including rescattering effects. The unpolarized and helicity-dependent total cross sections are calculated and compared with the available experimental data. In addition, results of total $\pi^0$-photoproduction cross sections from the free nucleon using the isobar MAID-2007 model are predicted. We obtained a satisfactory description of the experimental data from MAMI [2] for the total helicity-dependent cross section difference $\sigma^P - \sigma^A$. Furthermore, the contribution of coherent and incoherent $\pi^0$-photoproduction from the deuteron to the finite GDH integral has been evaluated by explicit integration up to 1.5 GeV and a total value of 256.96 μb has been obtained. Convergence of the GDH integral has been reached. Future theoretical improvements should be devoted to the inclusion of two-body effects in the $\pi^0$-photoproduction amplitudes.

## Acknowledgements

This work was supported in part by the Deanship of Scientific Research of the Taibah University, Saudi Arabia. We would like to thank E. Moya de Guerra, J.M. Udias, C. Fernandez-Ramirez, N. Akopov, and S. Al-Thoyaib for their valuable assistance. We are deeply indebted to J. Ahrens, P. Pedroni, and B. Krusche for providing us with their experimental data.

## References

[1] Krusche, B., Schadmand, S. 2003. Study of non-strange baryon resonances with meson photoproduction. Prog. Part. Nucl. Phys. 51: 399-485, http://dx.doi.org/10.1016/S0146-6410(03)90005-6.

[2] Burkert, V., Lee, T.-S. H. 2004. Electromagnetic meson production in the nucleon resonance region. Int. J. Mod. Phys. E 13: 1035, http://dx.doi.org/10.1142/S0218301304002545.

[3] Krusche, B., Ahrens, J., Beck, R., Fuchs, M., Hall, S.J., Harter, F., Kellie, J.D., Metag, V., Robig-Landau, M., Stroher, H. 1999. Single and double $\pi^0$-photoproduction from the deuteron. Eur. Phys. J. A. 6(3): 309-324, http://dx.doi.org/10.1007/s100500050349.

[4] Ilieva, Y., Berman, B.L., Kudryavtsev, A.E., Strakovsky, I.I., Tarasov, V.E., Amarian, M., et al. 2010. Evidence for a backward peak in the $\gamma d \rightarrow \pi^o d$ cross section near the $\eta$ threshold. Eur. Phys. J. A 43: 261-267, http://dx.doi.org/10.1140/epja/i2010-10918-x.

[5] Ahrens, J., Altieri, S., Annand, J.R., Arends, H.J., Beck, R., Bradtke, C., et al. 2006. Measurement of the Gerasimov-Drell-Hearn integrand for ²H from 200 to 800 MeV. Phys. Rev. Lett. 97(20): 202303, http://dx.doi.org/10.1103/PhysRevLett.97.202303.

[6] Ahrens, J., Altieri, S., Annand, J.R.M., Arends, H.-J., Beck, R., Blacjston, M.A., et al. 2009. Helicity dependence of the total inclusive cross section on the deuteron. Phys. Lett. B. 672(4-5): 328-332, http://dx.doi.org/10.1016/j.physletb.2009.01.061.

[7] Ahrens, J., Arends, H.J., Beck, R., Heid, E., Jahn, O., Lang, M., et al. 2020. Helicity dependence of the $\gamma d \rightarrow \pi NN$ reactions in the $\Delta$-resonance region. Eur. Phys. J. A 44: 189-201, http://dx.doi.org/10.1140/epja/i2010-10946-6.

[8] Darwish, E.M., Arenhovel, H., Schwamb, M. 2003. Influence of final-state interaction on incoherent pion photoproduction on the deuteron in the region of the $\Delta$-resonance. Eur. Phys. J. A 16: 111-125, http://dx.doi.org/10.1140/epja/i2002-10071-3.

[9] Darwish, E.M., Fernandez-Ramirez, C., Moya de Guerra, E., Udias, J.M. 2007. Helicity dependence and contribution to the Gerasimov-Drell-Hearn sum rule of the $\vec{\gamma} \vec{d} \rightarrow \pi NN$ reaction channels in the energy region from threshold up to the $\Delta(1232)$ resonance. Phys. Rev. C 76: 044005, http://dx.doi.org/10.1103/PhysRevC.76.044005.

[10] Darwish, E.M., Al-Thoyaib, S.S. 2011. Incoherent pion photoproduction on the deuteron including polarization effects. Ann. Phys. (N.Y.) 326(3): 604-625, http://dx.doi.org/10.1103/PhysRevC.76.044005.

[11] Arenhövel, H., Fix, A., Schwamb M. 2004. Spin asymmetry and Gerasimov-Drell-Hearn sum rule for the deuteron. Phys. Rev. Lett. 93(20): 202301, http://dx.doi.org/10.1103/PhysRevLett.93.202301.

[12] Schwamb, M. 2010. Unified description of hadronic and electromagnetic reactions of the two-nucleon system. Phys. Rep. 485(4-5): 109-193, http://dx.doi.org/10.1016/j.physrep.2009.10.002.

[13] Levchuk, M.I. Loginov, A.Yu., Sidorov, A.A., Stibunov, V.N., Schumacher, M. 2006. Incoherent pion photoproduction on the deuteron in the first resonance region. Phys. Rev. C 74: 014004, http://dx.doi.org/10.1103/PhysRevC.74.014004.

[14] Levchuk, M.I. 2010. Helicity-dependent reaction $\vec{\gamma}\vec{d} \rightarrow \pi NN$ and its contribution to the Gerasimov-Drell-Hearn sum rule for the deuteron. Phys. Rev. C

82: 044002, http://dx.doi.org/10.1103/PhysRevC.82.044002.

[15] Tarasov, V.E., Briscoe, W.J., Gao, H., Kudryavtsev, A.E., Strakovsky, I.I. 2011. Extracting the photoproduction cross sections off the neutron, via the γn→π⁻p reaction, from deuteron data with final-state interaction effects. Phys. Rev. C 84: 035203, http://dx.doi.org/10.1103/PhysRevC.84.035203.

[16] Arenhövel, H., Fix, A. 2005. Incoherent pion photoproduction on the deuteron with polarization observables. I. Formal expressions. Phys. Rev. C 72: 064004, http://dx.doi.org/10.1103/PhysRevC.72.064004.

[17] Fix, A., Arenhövel, H. 2005. Incoherent pion photoproduction on the deuteron with polarization observables. II. Influence of final state rescattering. Phys. Rev. C 72: 064005, http://dx.doi.org/10.1103/PhysRevC.72.064005.

[18] Darwish, E.M., Hemmdan, A. 2015. Influence of intermediate ηNN interaction on spin asymmetries for γd→π°d reaction near the η-threshold within a three-body approach. Ann. Phys. (N.Y.) 356: 128-148, http://dx.doi.org/10.1016/j.aop.2015.02.035.

[19] Darwish, E.M., Al-Thoyaib, S.S. 2015. Sensitivity of γd→π⁰d observables near η-threshold to the intermediate ηNN interaction and the choice of elementary pion production amplitude. J. Phys. Soc. Jpn. 84: 124201, http://dx.doi.org/10.7566/JPSJ.84.124201.

[20] Darwish, E.M., Hemmdan, A., El-Shamy, N.T. 2015. Helicity-dependent reaction γd → π°d near the η-threshold and its contribution to the E-asymmetry and the GDH sum rule for the deuteron. Int. J. Mod. Phys. E. 24: 1550064, http://dx.doi.org/10.1142/S0218301315500640.

[21] Fix, A., Arenhövel, H. 2002. The *ηNN*-system at low energy within a three-body approach. Nucl. Phys. A 697: 277-302, http://dx.doi.org/10.1016/S0375-9474(01)01230-1.

[22] Fix, A. 2005. Signature of the ηNN configurations in coherent π-photoproduction on the deuteron. Eur. Phys. J. A 26: 293-299, http://dx.doi.org/10.1140/epja/i2005-10167-2.

[23] Drechsel, D., Kamalov, S.S., Liator, L. 2007. Unitary isobar model -- MAID2007. Eur. Phys. J. A. 34: 69-97, http://dx.doi.org/10.1140/epja/i2007-10490-6.

[24] Gerasimov, S.B. 1965. A Sum rule for magnetic moments and the damping of the nucleon magnetic moment in nuclei. Yad. Fiz. 2: 598-602 [Sov. J. Nucl. Phys. 2 (1966) 430-433].

[25] Drell, S.D., Hearn, A.C. 1966. Exact sum rule for nucleon magnetic moments. Phys. Rev. Lett. 16: 908-911, http://dx.doi.org/10.1103/PhysRevLett.16.908.

[26] Darwish, E.M. 2002. Rescattering effects in incoherent photoproduction of π-mesons off deuterium in the Delta (1232) resonance region. PhD Thesis, University of Mainz, Germany, arXiv:nucl-th/0303056.

[27] Haidenbauer, J., Koike, Y., Plessas, W. 1986. Separable representation of the Bonn nucleon-nucleon potential. Phys. Rev. C. 33: 439, http://dx.doi.org/10.1103/PhysRevC.33.439.

[28] Dennery, P. 1961. Theory of the electro- and photoproduction of π mesons. Phys. Rev. 124: 2000, http://dx.doi.org/10.1103/PhysRev.124.2000.

[29] Siodlaczek, U., Achenbach, P., Ahrens, J., Annand J.R.M., Arends, H.-J., Beck, R., et al. 2001. Coherent and incoherent π°-photoproduction from the deuteron. Eur. Phys. J. A. 10: 365-371, http://dx.doi.org/10.1007/s100500170120.